\documentclass[11pt]{article}
\usepackage{amsmath, amsfonts, bm, amssymb}
\usepackage{amsfonts, graphicx, grffile, float, latexsym, mathtools}
\usepackage{dsfont}
\usepackage{multicol}
\usepackage{comment}
\usepackage{cite}
\usepackage{wrapfig}
\AtBeginEnvironment{algorithm}{\setlength{\intextsep}{6pt} \setlength{\textfloatsep}{6pt}}
\usepackage{tikz}
\usetikzlibrary{positioning}
\usetikzlibrary{calc}
\usepackage{wrapfig}
\usepackage{float}
\usepackage{mathrsfs}
\makeatletter
\let\NAT@parse\undefined
\makeatother
\usepackage{amsthm}
\usepackage{bbm}
\usepackage{bm}

\makeatletter
\renewcommand*\env@matrix[1][c]{\hskip -\arraycolsep
  \let\@ifnextchar\new@ifnextchar
  \array{*\c@MaxMatrixCols #1}}

\usepackage{Shorthands}
\usepackage{enumitem}

\makeatletter
\let\NAT@parse\undefined
\makeatother

\hypersetup{%
    pdfborder = {0 0 0}, citecolor=black
}

\usepackage{cleveref}
\crefname{assumption}{}{}
\newtheoremstyle{named}{}{}{\itshape}{}{\bfseries}{.}{.5em}{\thmnote{#3}}
\theoremstyle{named}

\newcommand{\ipar}{{(i)}}
\newcommand{\E}{\ensuremath{\mathbf{E}}}
\newcommand{\stab}{\textup{stab}}
\newcommand{\avg}{\textup{avg}}
\usepackage{mathrsfs}

\usepackage{algorithm} 
\usepackage{algpseudocode} 
\usepackage{authblk}

\usepackage{caption}
\renewcommand{\qed}{\hfill\blacksquare}

\usepackage{authblk}
\newcommand{\thickbar}[1]{\bm\bar{#1}}

\usepackage{fullpage}

\newcommand{\inner}[2]{\langle #1,\, #2 \rangle}

\title{Policy Gradient Bounds in Multitask LQR}

\author[1]{Charis Stamouli\thanks{Charis Stamouli and Leonardo F. Toso contributed equally.}}
\author[2]{Leonardo F. Toso$^*$}
\author[3]{Anastasios Tsiamis}
\author[1]{\\George J. Pappas}
\author[2]{James Anderson}

\affil[1]{University of Pennsylvania}
\affil[2]{Columbia University}
\affil[3]{ETH Zürich}

\date{}

\begin{document}

\maketitle
\thispagestyle{empty}
\pagestyle{empty}
\captionsetup[figure]{labelfont={bf},labelformat={default},labelsep=period,name={Fig.}}

\begin{abstract}
We analyze the performance of policy gradient in multitask linear quadratic regulation (LQR), where the system and cost parameters differ across tasks. The main goal of multitask LQR is to find a controller with satisfactory performance on every task. Prior analyses on  relevant contexts fail to capture closed-loop task similarities, resulting in conservative performance guarantees. To account for such similarities, we propose bisimulation-based measures of task heterogeneity. Our measures employ new bisimulation functions to bound the cost gradient distance between a pair of tasks in closed loop with a common stabilizing controller. Employing these measures, we derive suboptimality bounds for both the multitask optimal controller and the asymptotic policy gradient controller with respect to each of the tasks. We further provide conditions under which the policy gradient iterates remain stabilizing for every system. For multiple random sets of certain tasks, we observe that our bisimulation-based measures improve upon baseline measures of task heterogeneity dramatically.
\end{abstract}

\section{Introduction}
Designing a control policy that performs effectively on tasks with heterogeneous dynamics and objectives is a central problem in multitask reinforcement learning.
For a collection of $N$ tasks, a typical approach is to find a controller or policy parameter $K_\star$ that minimizes the average cost:
\begin{align}\label{eq:average_cost_function}
    J_{\avg}(K) := \frac{1}{N}\sum_{i=1}^{N} J^{(i)}(K)
\end{align}
over a parameter class containing $K_\star$. A popular method for approximating $J_{\avg}(K_\star)$ is provided by policy gradient \cite{teh2017distral}, with applications ranging from autonomous driving \cite{Qi2020} to robotic control \cite{kalashnikov2021mt}. Despite the empirical success of policy gradient, its theoretical guarantees remain relatively unexplored.

In this paper, we analyze the performance of vanilla policy gradient \cite{fazel2018global} in multitask linear quadratic regulation (LQR) with respect to each task. Our setting involves LQR tasks with heterogeneous system and cost parameters.  In this setting, closed-loop stability may not necessarily be preserved for every system across policy gradient iterations. Our analysis relies on closed-loop measures of task heterogeneity, inspired by classical bisimulation functions \cite{girard2005}. Bisimulation functions, as introduced in \cite{girard2005}, are Lyapunov-like functions that provide a principled method for characterizing the output distance of stable systems with vector states. Our contributions are the following:
\vspace*{-0.0cm}
\begin{itemize}
\item We define a novel notion of bisimulation functions tailored to Lyapunov matrix systems, and provide a systematic design of them via linear matrix inequalities.
\item Employing these functions, we introduce bisimulation-based measures that bound the cost gradient discrepancy between two LQR tasks under a common stabilizing controller. Our result establishes the first closed-loop measure of task heterogeneity for this purpose.
\item Employing these measures, we provide suboptimality bounds for the multitask optimal controller $K_\star$ and the asymptotic policy gradient controller with respect to each task. These bounds depend on the average bisimulation-based measure between the task of interest and the others, evaluated at the respective controller.  
\item We identify conditions such that the policy gradient iterates remain stabilizing for all systems.
\end{itemize}

We apply multitask policy gradient LQR across two sets of tasks: one with inverted pendula and another with unicycles. We observe that our bisimulation-based measures can be informative of the multitask policy gradient controller's performance in cases where previous measures from \cite{toso2024meta,toso2024asynchronous} are vacuous. 
For multiple random sets of these tasks, our measures dramatically improve upon baselines, effectively mitigating their conservatism. Complete proofs of our results are given in Appendix~\ref{app:Proofs}. The related work is summarized below. 
\vspace{0.1cm}

\noindent \textbf{Policy Gradient Methods.} Recent work has studied variants of our setting with different policy gradient methods. The authors of \cite{wang2023model,fujinami2025policy} consider the special case where the cost parameters are the same for all tasks. A related line of work focuses on meta-learning LQR \cite{toso2024meta}, which includes additional fine-tuning to each task-specific cost and presents meta-LQR design with heterogeneous systems and objectives. Closest to our setting is the work in \cite{toso2024asynchronous}, which proposes an asynchronous policy gradient approach for multitask LQR under diverse system and cost parameters. 

The analyses in both \cite{toso2024meta} and \cite{toso2024asynchronous} provide suboptimality bounds for policy gradient methods based on a measure of the heterogeneity between different tasks. Their measure bounds the norm of the cost gradient difference between the tasks via the maximum norm distance between the task parameters. The derivation of their measure overlooks potential closed-loop task similarities, thus often leading to overly conservative suboptimality bounds. In contrast, our bisimulation-based measures provide a principled \emph{closed-loop} notion of task heterogeneity, resulting in more informative performance bounds.  
\vspace{0.1cm}

\noindent \textbf{Bisimulations for Reinforcement Learning and Control.} Behavioral similarity has been central in both robust control (e.g., gap metric \cite{georgiou1989optimal,zhou1998essentials}) and layered control design (e.g., bisimulation relations \cite{haghverdi2003bisimulation,van2004equivalence} and functions \cite{girard2005,girard2011approximate}). Approximate bisimulation, in particular, employs Lyapunov-like functions to bound the output distance of two stable systems. Unlike our novel notion of bisimulations, classical bisimulations focus only on behavioral similarity with respect to system dynamics and remain agnostic to task objectives. The same limitation applies to the gap metric. 

In reinforcement learning, approximate bisimulation has been used to capture state similarity in Markov decision processes with respect to immediate rewards and the distribution of next states under a given policy \cite{castro2010using,ferns2014bisimulation,kemertas2022approximate}, and has been leveraged for policy transfer across similar states. These measures do not quantify deviations in policy gradient descent directions across tasks, where similarity must jointly capture both dynamics and objectives in the cost gradients. Our work addresses this gap by introducing a bisimulation-based measure that bounds the cost gradient discrepancy across LQR tasks under a common stabilizing controller.
\vspace{0.1cm}

\noindent\textbf{Notation.} The norm $\norm{\cdot}$ is the Euclidean norm when applied to vectors and the spectral norm when applied to matrices, while $\norm{\cdot}_F$ denotes the Frobenius norm. For a square matrix $A$, let $\trace{A}$, $\rho(A)$, $\lambda_{\min}(A)$, and $\sigma_{\min}(A)$  denote the trace, spectral radius, minimum eigenvalue, and minimum singular value, respectively. We use the notation $\mydiag(A,B)$ to denote the block-diagonal matrix with $A$ and $B$ on its diagonal, and $\mathds{I}_d$ to represent the $d \times d$ identity matrix. The notation $\setS_+^d$ refers to the set of positive definite $d \times d$ matrices. Expectation with respect to all the randomness of the underlying probability space is denoted by $\E$.

\section{Problem Formulation}

Consider a collection of $N$ discrete-time systems:
\begin{equation*}\label{eq:systems}
    x_{t+1}^{(i)} = A^\ipar x_t^{(i)}+B^{(i)} u_t^\ipar,\;i=1,\ldots,N, 
\end{equation*}
where $x_t^{(i)}\in\setR^{d_x}$ and $u_t^{(i)}\in\setR^{d_u}$ denote the state and control input of the $i$-th system, respectively. The initial states $x_0^{(i)}$ are zero-mean random variables with covariance $\Sigma_0^{(i)}\succ0$. The performance of system $i$ under the feedback policy $u_t^{(i)}=-Kx_t^{(i)}$, with $K\in\setR^{d_u\times d_x}$, is measured by the quadratic cost: 
\begin{equation*}\label{eq:task_specific_cost_functions}
    J^{(i)}(K):= \E_{}\left[\sum_{t=0}^{\infty}x_t^{(i)\intercal}\left(Q^{(i)}+K^{\intercal}R^{(i)}K\right)x_t^{(i)}\right],
\end{equation*}
where $Q^{(i)}\succeq0$ and $R^{(i)}\succ0$. Consider the set of common stabilizing controllers
$\calK_\stab:=\cap_{i=1}^N\calK_\stab^{(i)},$
where $\calK_\stab^{(i)}$ is the set of stabilizing controllers for system $i$. For each $i$, we define the $i$-th LQR task as $\mathcal{T}^{(i)} = \big(A^{(i)}, B^{(i)}, Q^{(i)}, R^{(i)}\big)$  
and its optimal controller as $K_\star^{(i)}=\argmin_{K\in \calK_{\stab}^{(i)}} J^{(i)}(K)$. The objective of multitask LQR is to find $K_\star\in\calK_\stab$ that minimizes the average cost over all tasks, given by \eqref{eq:average_cost_function}.

In multitask reinforcement learning, a standard method of approximating the minimum value $J_\avg(K_\star)$ is that of policy gradient \cite{teh2017distral}. In our setting, we apply policy gradient with an initial controller $K_0\in\calK_\stab$\footnote{A controller $K_0\in\calK_\stab$ can be efficiently computed \cite{fujinami2025policy}.} and iterative updates:
\begin{align}\label{eq:policy_gradient}
    K_{n+1}=K_n-\alpha\nabla J_\avg(K_n),
\end{align}
where $\alpha\in\setR_+$ is a fixed step size. For each $i$ and any 
controller $K\in\calK_\stab$, we define the dynamics $x_{K,t+1}^{(i)}=A_K^{(i)}x_{K,t}^{(i)}$, where $A_K^{(i)}=A^{(i)}-B^{(i)}K$, and $P_K^{(i)}$ as the unique solution of the Lyapunov equation: 
\begin{align*}
    P_K^{(i)} - A_K^{(i)\intercal} P_K^{(i)}A_K^{(i)}=Q^{(i)}+K^\intercal R^{(i)}K.
\end{align*}
From \cite[Lemma 1]{fazel2018global} it follows that for any $K\in\calK_\stab$, the gradient of the average cost can be expressed as:
\begin{equation}\label{eq:avg_cost_gradient}
    \nabla J_\avg(K) = \frac{1}{N}\sum_{i=1}^N\nabla J^{(i)}(K)=\frac{1}{N}\sum_{i=1}^NE_K^{(i)}\Sigma_K^{(i)},
\end{equation}
where for each $i$:
\begin{align}
\label{eq:E_K}
&E_K^{(i)}=2((R^{(i)}+B^{(i)\intercal}P_K^{(i)}B^{(i)})K-B^{(i)\intercal}P_K^{(i)}A^{(i)}),\\
&\Sigma_K^{(i)}=\E\left[\sum_{t=0}^\infty x_{K,t}^{(i)}x_{K,t}^{(i)\intercal}\right].
\end{align}

Our paper aims to analyze the task-specific performance of $K_\star$ and  $K_n$ with respect to that of each $K_\star^{(i)}$. More specifically, we are interested in providing upper bounds for:
\begin{enumerate}[label=\roman*)]
\item The task-specific optimality gaps of $K_\star$:
\begin{align}
    \label{eq:task_specific_gap_K_star}
    &J^{(i)}(K_\star)-J^{(i)}(K_\star^{(i)}),\;i=1,\ldots,N.
\end{align}
\item The task-specific asymptotic optimality gaps of $K_n$:
\begin{align}
    \label{eq:task_specific_gap_K_infty}
    \limsup_{n\to\infty}(&J^{(i)}(K_n)-J^{(i)}(K_\star^{(i)})),\;i=1,\ldots,N.
\end{align}
\end{enumerate}

In similar settings, prior work \cite{toso2024meta,toso2024asynchronous} has 
established the gradient gaps between pairs of tasks  $(i,j)$:
\begin{equation}\label{eq:cost_grad_gap}
g_{ij}(K):=\norm{\nabla J^{(i)}(K)-\nabla J^{(j)}(K)}_F
\end{equation}
as fundamental for bounding the task-specific optimality gaps. Therefore, we say that $g_{ij}(K)$ quantifies the \emph{heterogeneity} between tasks $\calT^{(i)}$ and $\calT^{(j)}$ under any $K\in\calK_\stab$.

\subsection{Limitations of Existing Task Heterogeneity Measures}\label{subsec:Limitations of Open-Loop Task Heterogeneity Measures}
To measure task heterogeneity, prior work \cite{toso2024meta,toso2024asynchronous} begins by assuming bounded deviations in the system parameters:
\begin{align*}
    \label{eq:b_A_b_B}
    &\max_{i\neq j} \norm{A^{(i)}-A^{(j)}}\leq b_A, ~~ \max_{i\neq j} \norm{B^{(i)}-B^{(j)}}\leq b_B.
\end{align*}
Bounds $b_Q$ and $b_R$ for the cost parameter deviations are similarly defined. The above deviation bounds are combined into a task heterogeneity measure, $\bar{b}(K) := \bar{b}(K;b_A,b_B,b_Q,b_R)$, which for any $K\in\calK_\stab$ satisfies:
\begin{equation}\label{eq:prev_cost_grad_het_bound}
\max_{i\neq j}g_{ij}(K)\leq\bar{b}(K).
\end{equation}
For the exact expression of $\bar{b}(K)$, we refer the reader to \cite[Appendix 6.2]{toso2024asynchronous}. The measures $\bar{b}(K_\star)$ and $\bar{b}(K_n)$ are used to bound the performance gaps $J^{(i)}(K_\star)-J^{(i)}(K_\star^{(i)})$ and $J^{(i)}(K_n)-J^{(i)}(K_\star^{(i)})$, respectively. However, they  are often vacuous in practice, even when the actual performance gaps are small, as illustrated in the following example.

\begin{example}[Inverted Pendulum]\label{example}
Consider a collection of\; $6$ LQR tasks corresponding to the linearized inverted pendulum dynamics  with system matrices:
\begin{equation*}
    A^{(i)} = \begin{bmatrix}
    1 & dt \\
    \frac{g}{\ell_i}\,dt & 1
    \end{bmatrix},\; 
    B^{(i)} = \begin{bmatrix}
    0 \\
    \frac{dt}{m_i \ell_i^{2}}
\end{bmatrix},
\end{equation*}
where $\ell_i$ is the length, $m_i$ is the mass, $g=10$, and $dt=0.05$. The LQR cost matrices are $Q^{(i)}=q_i\mathds{I}_2$ and $R^{(i)}=r_i$, and the initial state covariances are  $\Sigma_0^{(i)}=0.01\mathds{I}_2$. The parameters $\ell_i$, $m_i$, $q_i$, and $r_i$ are uniformly sampled from $[0.5,1]$, $[0.1,0.5]$, $[0.1,0.5]$, and $[0.01,0.05]$, respectively. As shown in Fig.~\ref{fig:inverted_pendulum_cost_gap_GOOD}, the optimality gaps \eqref{eq:task_specific_gap_K_infty} are below $0.15$ at convergence, for a step size of $0.01$, indicating satisfactory performance of the multitask policy gradient controller. This is in contrast to the heterogeneity measure $\bar{b}(K_n)$, which evaluates to $2.3 \times 10^6$ at convergence. Its conservatism can be explained by the derivation of  $\bar{b}(K)$ in terms of parameter deviation bounds. Specifically, despite the dependence of $\bar{b}(K)$ on $K$, its construction overlooks potential closed-loop task similarities that could lead to smaller task heterogeneity.  

\begin{figure}[H]
      \centering
      %\framebox{\parbox{3in}{}}}
      \includegraphics[width=.53\linewidth]{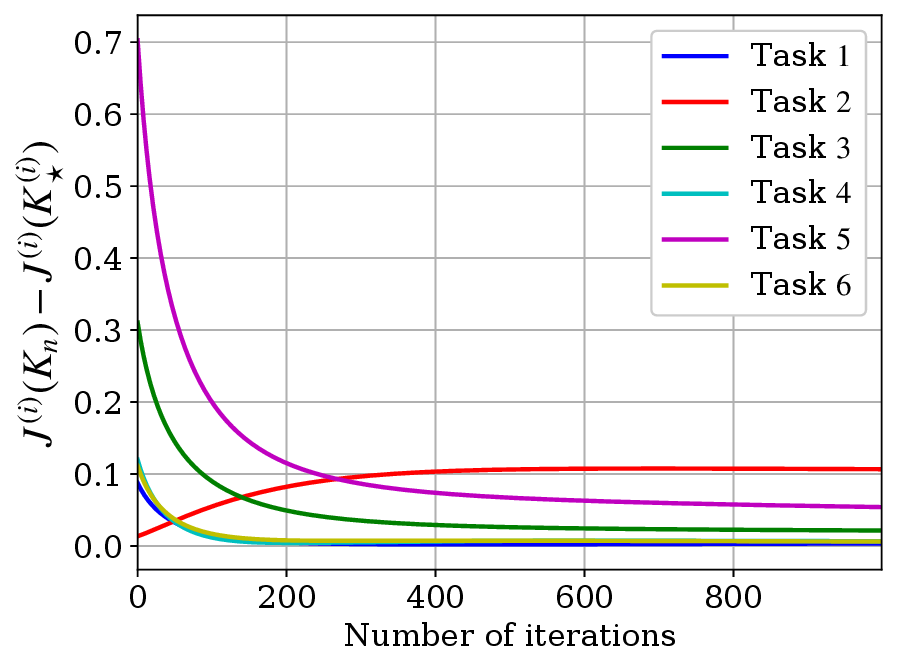}
      \vspace{-0.3cm}
      \caption{Evolution of optimality gaps associated with the multitask policy gradient iterates for $6$ inverted-pendulum tasks.}
      \label{fig:inverted_pendulum_cost_gap_GOOD}
\end{figure}
\end{example}

\subsection{Toward Closed-Loop Task Heterogeneity Measures}

Our goal is to develop closed-loop measures of task heterogeneity, suitable for bounding the performance of both the multitask optimal controller and the policy gradient iterates. Exploiting the closed-loop behavior is intended to mitigate the conservatism incurred by the previous use of parameter deviation bounds to analyze the gaps \eqref{eq:task_specific_gap_K_star} and \eqref{eq:task_specific_gap_K_infty}. 

Toward this end, we next present a novel notion of bisimulation-based measures of task heterogeneity, which bound the gradient gaps in \eqref{eq:cost_grad_gap}. In Section~\ref{sec:Bisimulation-Based Performance Analysis of Multitask LQR}, we leverage these measures to bound the optimality gaps \eqref{eq:task_specific_gap_K_star} and \eqref{eq:task_specific_gap_K_infty}.

\section{Bisimulation-Based Task Heterogeneity}\label{sec:Bisimulation-Based Task Heterogeneity}
In this section, we propose bisimulation-based measures that allow us to  bound the discrepancy in cost gradients between tasks, as quantified by \eqref{eq:cost_grad_gap}. Unlike existing measures of task heterogeneity in \cite{toso2024meta,toso2024asynchronous}, our measures provide a \emph{closed-loop}
notion of heterogeneity between different tasks. In  Subsection~\ref{subsec:Task Heterogeneity based on Bisimulation Functions}, we define bisimulation  
functions that can be used to bound the gradient discrepancies \eqref{eq:cost_grad_gap}. In Subsection~\ref{subsec:Characterization of Bisimulation Functions}, we develop an effective derivation of these functions and formalize our bisimulation-based measures. 

\subsection{Bisimulation Functions for Cost Gradient Discrepancies}\label{subsec:Task Heterogeneity based on Bisimulation Functions}
We now introduce novel bisimulation functions that allow us to analyze the gradient gaps \eqref{eq:cost_grad_gap}. 
For every task $\calT^{(i)}$ and any  $K\in\calK_\stab$, we consider the associated matrix system:
\begin{align}\label{eq:matrix_system}
    S_K^{(i)}:\left\{
                \begin{array}{lcr}
                    \Sigma_{K,t+1}^{(i)}=A_K^{(i)} \Sigma_{K,t}^{(i)}A_K^{(i)\intercal}+\Sigma_0^{(i)}\\
                    Y_{K,t}^{(i)}=E_K^{(i)} \Sigma_{K,t}^{(i)}
                \end{array}
                \right.,
\end{align}
where $\Sigma_{K,t}^{(i)}\in\setS_+^{d_x}$ and $E_K^{(i)}$ is defined by \eqref{eq:E_K}. The dynamics in \eqref{eq:matrix_system} represent those of system $i$'s covariance under controller $K$. By using \cite[Lemma 1]{fazel2018global} and the fact that $K\in\calK_\stab$, we can deduce that 
the output $Y_{K,t}^{(i)}$ converges to the cost gradient $\nabla J^{(i)} (K)$. Therefore, we have:
\begin{equation}\label{eq:grad_het_limit}
    \norm{\nabla J^{(i)}(K)-\nabla J^{(j)}(K)}_F=\lim_{t\to\infty}\norm{Y_{K,t}^{(i)}-Y_{K,t}^{(j)}}_F
\end{equation}
and can focus on bounding the asymptotic output distance of systems $S_K^{(i)}$ and $S_K^{(j)}$.

The concept of bisimulation functions, as introduced in \cite{girard2005}, provides a principled method for characterizing the output distance between two  stable continuous-time systems with vector states. Next, we extend the definition of bisimulation functions to affine matrix systems of the form \eqref{eq:matrix_system}. 

\begin{definition}[Bisimulation Functions for Lyapunov Matrix Systems]\label{def:bisim_function}
A continuous function $V:\setS_+^{d_x}\times \setS_+^{d_x}\to\setR_+$ is a bisimulation function between $S_K^{(i)}$ and $S_K^{(j)}$ if there exists $\lambda\in(0,1)$ such that for all
$(\Sigma_{K,t}^{(i)},\Sigma_{K,t}^{(j)})\in\setS_+^{d_x}\times \setS_+^{d_x}$:
\begin{subequations}
\begin{align}
\label{eq:BF_condition_1}
    V&(\Sigma_{K,t}^{(i)},\Sigma_{K,t}^{(j)})\geq \norm{E_K^{(i)} \Sigma_{K,t}^{(i)}-E_K^{(j)} \Sigma_{K,t}^{(j)}}_F, 
\end{align}
\begin{align}
\label{eq:BF_condition_2}
    V(\Sigma_{K,t+1}^{(i)},&\Sigma_{K,t+1}^{(j)})-V(\Sigma_{K,t}^{(i)},\Sigma_{K,t}^{(j)})\leq -\lambda V(\Sigma_{K,t}^{(i)},\Sigma_{K,t}^{(j)})+V(\Sigma_0^{(i)},\Sigma_0^{(j)}).
\end{align}
\end{subequations}
\end{definition}

Condition \eqref{eq:BF_condition_1} implies that a bisimulation function $V(\cdot,\cdot)$ provides an upper bound for the output distance of systems $S_K^{(i)}$ and $S_K^{(j)}$, given any pair of  states. Condition \eqref{eq:BF_condition_2} can be easily shown to guarantee that $V(\Sigma_{K,t}^{(i)},\Sigma_{K,t}^{(j)})$ is asymptotically upper-bounded by $V(\Sigma_0^{(i)},\Sigma_0^{(j)})/\lambda$. 

\begin{lemma}\label{lem:asympt_V_bound}
Let $V:\setS_+^{d_x}\times \setS_+^{d_x}\to\setR_+$ denote a bisimulation function between $S_K^{(i)}$ and $S_K^{(j)}$. It holds that:
\begin{equation}\label{eq:asympt_bis_function_bound}
    \lim_{t\to\infty}\norm{Y_{K,t}^{(i)}-Y_{K,t}^{(j)}}_F\leq V(\Sigma_0^{(i)},\Sigma_0^{(j)})/\lambda. 
\end{equation}
\end{lemma}

The proof follows from condition \eqref{eq:BF_condition_1} and a recursive expansion of \eqref{eq:BF_condition_2}. Combining \eqref{eq:grad_het_limit} and \eqref{eq:asympt_bis_function_bound}, we obtain:
\begin{equation}\label{eq:grad_het_V_bound}
    \norm{\nabla J^{(i)}(K)-\nabla J^{(j)}(K)}_F\leq V(\Sigma_0^{(i)},\Sigma_0^{(j)})/\lambda. 
\end{equation}

Next, we show that for any $K \in \calK_\stab$, a bisimulation function $V(\cdot,\cdot)$ can be effectively designed for $S_K^{(i)}$ and $S_K^{(j)}$.

\subsection{Task Heterogeneity via Bisimulation Function Design}\label{subsec:Characterization of Bisimulation Functions}

In this subsection, we develop a systematic procedure for designing bisimulation functions between systems $S_K^{(i)}$ and $S_K^{(j)}$. Then, we leverage these functions to introduce bisimulation-based measures of task heterogeneity.

\begin{lemma}\label{lem:M_K_conditions}
Consider a controller $K\in\calK_\stab$ and a pair of task indices $(i,j)$.
There exist $M\in\setS_+^{2d_x}$ and $\lambda\in(0,1)$ such that the following linear matrix
inequalities (LMIs) hold:
\begin{subequations}
\begin{align}
    \label{eq:M_condition}
    &M\succeq E_K^{(ij)\intercal}E_K^{(ij)}, \\
    \label{eq:constraint_stabilization}
    &A_K^{(ij)\intercal}MA_K^{(ij)}-M\preceq-\lambda M,
\end{align}
\end{subequations}
where $E_K^{(ij)}=[E_K^{(i)} \; -E_K^{(j)}]$ and $A_K^{(ij)}=\mydiag(A_K^{(i)},A_K^{(j)})$.
\end{lemma}
The proof is similar to that of \cite[Lemma 2]{stamouli2025layered}. We next use the  linear matrix inequalities above to systematically characterize bisimulation functions.

\begin{theorem}[Bisimulation Functions for Lyapunov Matrix Systems]\label{theorem:Bisimulation_Functions_characterization}
Consider a controller $K\in\calK_\stab$ and a pair of task indices $(i,j)$. Let $M\in\setS_+^{2d_x}$ and $\lambda\in(0,1)$ satisfy \eqref{eq:M_condition} and \eqref{eq:constraint_stabilization}. The function $V:\setS_+^{d_x}\times \setS_+^{d_x}\to\setR_+$ defined by:
\begin{equation}\label{eq:bisimulation_function}
    V(\Sigma_{K,t}^{(i)},\Sigma_{K,t}^{(j)}) = \frac{\sqrt{2}\trace{M\mydiag(\Sigma_{K,t}^{(i)},\Sigma_{K,t}^{(j)})}}{\sqrt{\lambda_{\min}(M)}},
\end{equation}
is a bisimulation function between systems $S_K^{(i)}$ and $S_K^{(j)}$.
\end{theorem}

Combining \eqref{eq:grad_het_V_bound} and Theorem~\ref{theorem:Bisimulation_Functions_characterization} we can deduce that:
\begin{equation}\label{eq:grad_het_bis_bound}
    \norm{\nabla J^{(i)}(K)-\nabla J^{(j)}(K)}_F\leq \frac{\sqrt{2}\trace{M\mydiag(\Sigma_0^{(i)},\Sigma_0^{(j)})}}{\lambda \sqrt{\lambda_{\min}(M)}},
\end{equation}
for any $M\in\setS_+^{2d_x}$ and $\lambda\in(0,1)$ satisfying \eqref{eq:M_condition} and \eqref{eq:constraint_stabilization}.

\begin{definition}[Bisimulation-Based Task Heterogeneity]\label{def:bis_task_het}
Consider a controller $K\in\calK_\stab$ and a pair  of distinct task indices $(i,j)$. Let $M_K^{(ij)}$ and  $\lambda_K^{(ij)}$ denote the minimizers of the bound in \eqref{eq:grad_het_bis_bound} that satisfy the conditions of Lemma~\ref{lem:M_K_conditions}. The bisimulation-based heterogeneity between tasks $\calT^{(i)}$ and $\calT^{(j)}$, under the policy $u_t^{(i)}=-Kx_t^{(i)}$, is given by:
\begin{equation}\label{eq:bisim_het}
    b_{ij}(K) := \frac{\sqrt{2}\trace{M_K^{(ij)}\mydiag(\Sigma_0^{(i)},\Sigma_0^{(j)})}}{\lambda_K^{(ij)} \sqrt{\lambda_{\min}(M_K^{(ij)})}}.
\end{equation}
\end{definition}

The measure \eqref{eq:bisim_het} provides a closed-loop notion of task heterogeneity, as the LMI \eqref{eq:constraint_stabilization} considers  systems $i$ and $j$ in closed loop with the controller $K$. The quantities $M_K^{(ij)}$ and $\lambda_K^{(ij)}$ in \eqref{eq:bisim_het} can be efficiently computed via a convex optimization problem (see Appendix~\ref{app:Computation of Bisimulation Measures of Task Heterogeneity}).

\begin{remark}
The conditions in~\Cref{lem:M_K_conditions} are the discrete-time analogues of those derived for continuous-time linear systems with vector states in \cite{girard2005}. The bisimulation-based task heterogeneity  \eqref{eq:bisim_het} is the minimum bisimulation bound for the difference in the cost gradient responses of the $i$-th and $j$-th task, under a common $K\in\calK_\stab$, as in \eqref{eq:grad_het_bis_bound}. The value of $\lambda_K^{(ij)}$ is determined by the stability margin of $A_K^{(ij)}$; specifically, the larger the stability margin, the larger $\lambda_K^{(ij)}$ becomes (see  \eqref{eq:constraint_stabilization}). Moreover, $M_K^{(ij)}$ corresponds to the solution of a Lyapunov equation for $A_K^{(ij)}$, which dominates a matrix related to $E_K^{(i)}$ and $E_K^{(j)}$ (see \eqref{eq:M_condition} and \eqref{eq:constraint_stabilization}). These matrices are zero when  $K$ is equal to $K_\star^{(i)}$ and $K_\star^{(j)}$, respectively. Note also the dependence of $b_{ij}(K)$ on the initial state covariances $\Sigma_0^{(i)}$ and $\Sigma_0^{(j)}$, which directly affect the cost of the respective LQR tasks \cite{anderson2007optimal}. Subsequently, we demonstrate how the measures $b_{ij}(K)$ influence the optimality gaps  \eqref{eq:task_specific_gap_K_star} and \eqref{eq:task_specific_gap_K_infty}, while  reducing the conservatism of the previous task heterogeneity bounds \eqref{eq:prev_cost_grad_het_bound}.
\end{remark}

\section{Policy Gradient Bounds in Multitask LQR via Bisimulations}\label{sec:Bisimulation-Based Performance Analysis of Multitask LQR}
We now analyze the task-specific optimality gaps \eqref{eq:task_specific_gap_K_star} and \eqref{eq:task_specific_gap_K_infty} based on the bisimulation-based task heterogeneity measures \eqref{eq:bisim_het}. 

Before introducing our main results, we present a few definitions that will be needed for their statement. For every task $\calT^{(i)}$ and any $K\in\calK_{\stab}$, we define $b_i(K)$ as:
\begin{equation}\label{eq:avg_bis_het}
    b_i(K)=\frac{1}{N}\sum_{j\neq i} b_{ij}(K),
\end{equation}
where $b_{ij}(K)$ are given by \eqref{eq:bisim_het}. We note that $b_i(K)$ describes the average bisimulation-based heterogeneity between the task $\calT^{(i)}$ and each of the other tasks. We proceed with defining the stabilizing subset $\calK\subset\setR^{d_u\times d_x}$.

\begin{definition}[Stabilizing Subset]\label{def:stabilizing_set}
\cite{toso2024meta} 
Given a controller $K_0\in\calK_\stab$ and constants $\beta_1,\ldots,\beta_N\geq1$, the stabilizing subset is defined as $\calK:=\cap_{i=1}^N\calK^{(i)}$, where:
\begin{align*}
    \calK^{(i)}=\Big\{K\in\setR^{d_u\times d_x}&: J^{(i)}(K)-J^{(i)}(K_\star^{(i)})\leq\beta_i(J^{(i)}(K_0)-J^{(i)}(K_\star^{(i)}))\Big\}.
\end{align*}     
\end{definition}
A few observations about the set $\calK$ are in order. First, recall that $\calK_\stab$ is the set of common stabilizing controllers for all systems and note that $\calK\subseteq\calK_\stab$, since the cost of each task is finite for any controller in $\calK$. Second, observe the implicit dependence of $\calK$ on the given initial policy gradient controller $K_0$. Third, note that we can always pick constants $\beta_i$ large enough such that $K_\star\in\calK$. Therefore, the use of $\calK$ instead of $\calK_\stab$ does not limit our results, but rather simplifies our analysis, due to the compactness of $\calK$ shown in \cite{hu2023toward}. Specifically, from \cite[Theorem 1]{hu2023toward} it follows that the functions $J^{(i)}(\cdot)$ are smooth on the compact set $\calK$. The use of $\calK$ is also standard in prior work \cite{toso2024meta,toso2024asynchronous}.

\begin{theorem}[Bisimulation-Based Suboptimality Bounds for Multitask Optimal LQR]\label{thm:algor_indep_bounds}
Let $K_\star\in\calK_\stab$ be a minimizer of \eqref{eq:average_cost_function}. For every task $\calT^{(i)}$ and any  $K\in\calK_\stab$, let $b_{i}(K)$ be given by \eqref{eq:avg_bis_het}. Then, for each task index $i$, the task-specific optimality gap associated with  $K_\star$ is bounded as:
\begin{equation}\label{eq:gap_bound_K_star}
    J^{(i)}(K_\star)-J^{(i)}(K_\star^{(i)})\leq \frac{2\big\|\Sigma_{K_\star^{(i)}}^{(i)}\big\|\,b_i(K_\star)^2}{\lambda_{\min}(\Sigma_0^{(i)})^2\,\sigma_{\min}(R^{(i)})}.
\end{equation}
\end{theorem}

The proof results from adapting that of \cite[Theorem 3]{toso2024meta} to the setting of multitask LQR. We point out the dependence of the bound in \eqref{eq:gap_bound_K_star} on the squared bisimulation-based heterogeneity measure \eqref{eq:avg_bis_het}, evaluated at $K_\star$.

\begin{theorem}[Bisimulation-Based Suboptimality Asymptotics for Multitask Policy Gradient LQR]\label{thm:policy gradient_bounds}
Let $K_0\in\calK_\stab$  and consider the sequence $\{K_n\}_{n=0}^\infty$ of policy gradient iterates given by \eqref{eq:policy_gradient}. Consider $\beta_1,\ldots,\beta_N\geq1$, and let $\calK$ denote the stabilizing subset from~\Cref{def:stabilizing_set}. For each $i$, let $L_i$ denote the smoothness constant of the $i$-th LQR cost function $J^{(i)}(\cdot)$ over the set $\calK$. For every task $\calT^{(i)}$ and any  $K\in\calK_\stab$, let $b_{i}(K)$ be given by \eqref{eq:avg_bis_het}.  Assume that: $$\textstyle \alpha<\min\left\{1/(4\max_i L_i),4/\max_i\gamma_i\right\},$$ where $\gamma_i=(4\lambda_{\min}(\Sigma_0^{(i)})^2\sigma_{\min}(R^{(i)}))/\norm{\Sigma_{K_\star^{(i)}}^{(i)}}$, for all $i=1,\ldots,N$. Moreover, assume that: $$\sup_{K\in\calK}b_i(K)^2\leq \gamma_i(J^{(i)}(K_0)-J^{(i)}(K_\star^{(i)}))/6.$$ Then, we have $K_n\in\calK$, for all $n\in\setN$. Moreover, for each task index $i$, the task-specific asymptotic optimality gap associated with $\{K_n\}_{n=0}^\infty$ is bounded as:
\begin{align}\label{eq:gap_bound_K_infty} 
    \limsup_{n\to\infty}&(J^{(i)}(K_n)-J^{(i)}(K_\star^{(i)}))\leq \limsup_{n\to\infty}\frac{3\big\|\Sigma_{K_\star^{(i)}}^{(i)}\big\|\,b_i(K_n)^2}{4\lambda_{\min}(\Sigma_0^{(i)})^2\,\sigma_{\min}(R^{(i)})}.
\end{align}
\end{theorem}

The proof results from combining a modified version of the proof of \cite[Theorem 3]{toso2024meta} with a novel extension of \cite[Theorem 1]{hu2023toward} to the setting of multitask LQR. The conditions of~\Cref{thm:policy gradient_bounds} on a small enough step size and a low task heterogeneity compared to the initial optimality gap  $J^{(i)}(K_0)-J^{(i)}(K_\star^{(i)})$ are reasonable and standard in related work \cite{toso2024meta,toso2024asynchronous}. We highlight, though, that the use of bisimulation-based measures of task heterogeneity is novel.  We also note the dependence of the bound in \eqref{eq:gap_bound_K_infty} on the squared bisimulation-based heterogeneity measure \eqref{eq:avg_bis_het}, evaluated at the limit superior of policy gradient iterations. If the policy gradient iterates converge to some controller $K_{\infty}$, inequality \eqref{eq:gap_bound_K_infty} implies that:
\begin{align}\label{eq:gap_bound_K_infty_converges}
   J^{(i)}(K_\infty)-J^{(i)}(K_\star^{(i)})\leq \frac{3\big\|\Sigma_{K_\star^{(i)}}^{(i)}\big\|\,b_i(K_\infty)^2}{4\lambda_{\min}(\Sigma_0^{(i)})^2\,\sigma_{\min}(R^{(i)})},
\end{align}   
for all $i=1,\ldots,N$.

\begin{remark}[Result Interpretation]
The suboptimality bounds in \eqref{eq:gap_bound_K_star} and \eqref{eq:gap_bound_K_infty_converges} are equal up to universal constants and the bisimulation-based measures $b_i(K)$, which are evaluated at $K_\star$ and $K_\infty$, respectively. Their similarity is notable as the analysis of~\Cref{thm:algor_indep_bounds} is algorithm-independent and corresponds to the multitask optimal controller, while the analysis of~\Cref{thm:policy gradient_bounds} relies on the policy gradient algorithm defined by \eqref{eq:policy_gradient}. Beyond the  measures $b_i(K)$, the optimality gaps are bounded in terms of: i) the steady-state covariance of system $i$ under the task-specific optimal controller $K_\star^{(i)}$, ii) the initial state covariance $\Sigma_0^{(i)}$, and iii) the input cost matrix $R^{(i)}$.
\end{remark}

In the next section, we observe that our closed-loop measures \eqref{eq:avg_bis_het} significantly reduce the conservatism of previous heterogeneity measures and better capture the performance of multitask policy gradient LQR.

\section{Numerical Validation}
We next present numerical examples\footnote{Code for reproduction can be found at our \href{https://github.com/jd-anderson/multitask_bisim}{\textcolor{blue}{GitHub repository}}.} on the inverted pendulum and unicycle dynamics applying multitask policy gradient LQR to demonstrate the effectiveness of our bounds. We observe that our bisimulation-based measures are less conservative than the baseline bounds designed based on \cite{toso2024meta,toso2024asynchronous}, which bound the cost gradient discrepancies \eqref{eq:cost_grad_gap} in terms of the task parameter deviation bounds $b_A, b_B, b_Q,$ and $b_R$.\vspace{0.1cm}

\noindent \textbf{Inverted Pendulum.} We revisit  Example~\ref{example} and compare our bisimulation-based measures \eqref{eq:avg_bis_het} with baseline task heterogeneity measures $\bar{b}(K):=\bar{b}(K;b_A,b_B,b_Q,b_R)$ obtained from \eqref{eq:prev_cost_grad_het_bound}.

\begin{figure}[thpb]
      \centering
      \includegraphics[width=0.53\linewidth]{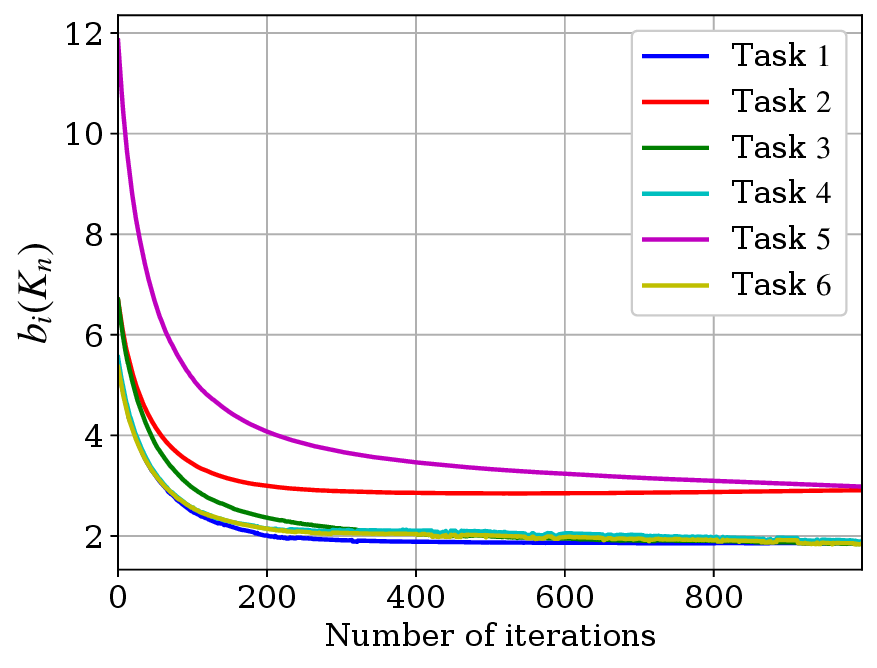}
      \vspace{-0.3cm}
      \caption{Evolution of bisimulation-based measures from \eqref{eq:avg_bis_het} in multitask policy gradient LQR for $6$ inverted-pendulum tasks.}
      \label{fig:bisimulation_bounds_inverted_pendulum}
\end{figure}

Suboptimality bounds similar to those in \eqref{eq:gap_bound_K_star} and \eqref{eq:gap_bound_K_infty_converges} can be derived in terms of $\bar{b}(K_\star)$ and $\bar{b}(K_\infty)$, respectively (see Subsection~\ref{subsec:Limitations of Open-Loop Task Heterogeneity Measures}). Recall that $\bar{b}(K_\infty)$ evaluates to $2.3\times 10^6$ in this example, despite the small optimality gaps observed in Fig.~\ref{fig:bisimulation_bounds_inverted_pendulum}. In contrast, our bisimulation-based measures depicted in Fig.~\ref{fig:bisimulation_bounds_inverted_pendulum} are below $3$ at convergence, reflecting the favorable task-specific performance of the multitask controller.

We repeat the computation of both measures over $100$ collections of inverted-pendulum tasks, which are generated as in Example~\ref{example}. In this case, we observe an average reduction of $99.9998\%$, which suggests a substantial decrease of conservatism with respect to the previous bounds. 

\noindent \textbf{Unicycle.} We consider a collection of $6$ LQR tasks corresponding to the linearized unicycle dynamics. The continuous-time kinematics are given by:
$$ \dot p_x = v \cos\theta,  \dot p_y = v \sin\theta, \text{ and }\dot \theta = \omega,$$
with state $x = (p_x, p_y, \theta)$ and input $u = (v, \omega)$, where $(p_x,p_y)$ is the robot's position on the $x,y$-plane, $\theta$ is the orientation angle, $v$ is the forward velocity, and $\omega$ is the yaw rate. 
We linearize the dynamics at operating points $(v_{0,i},\theta_{0,i})$ and then discretize with step size $dt=0.05$ to obtain the system matrices:
$$
A^{(i)}= 
\begin{bmatrix}
1 & 0 & -d_tv_{0,i} \sin\theta_{0,i} \\
0 & 1 & d_t v_{0,i} \cos\theta_{0,i} \\
0 & 0 & 1
\end{bmatrix}, \;
B^{(i)} = 
\begin{bmatrix}
d_t\cos\theta_{0,i} & 0 \\
d_t\sin\theta_{0,i} & 0 \\
0 & d_t
\end{bmatrix}.
$$
We uniformly draw $v_{0,i},\theta_{0,i},q_i$, and $r_i$ from the intervals $[0.1,1.75]$, $[0,\pi/2]$, $[0.1,0.5]$, $[0.01,0.05]$, respectively. We also set  $Q^{(i)} = q_i \mathds{I}_3$, $R^{(i)} = r_i \mathds{I}_2$, and initial state covariances $\Sigma_0^{(i)}=\mathds{I}_3,$ for all systems $i=1,\ldots,N$.

\begin{figure*}[thpb]
  \centering
  \begin{minipage}{.32\textwidth}
    \centering
    \includegraphics[width=1.\textwidth]{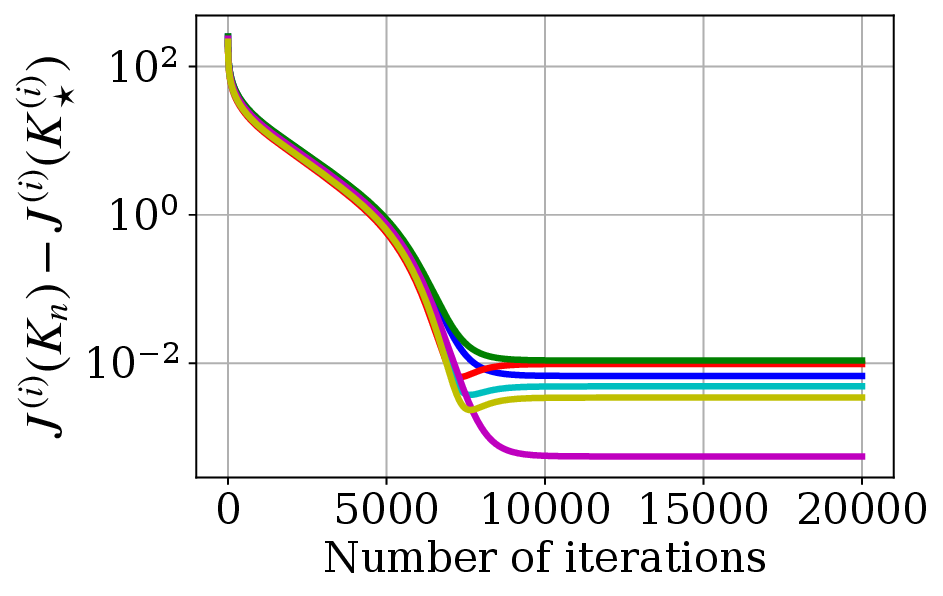}
  \end{minipage}
  \begin{minipage}{.3\textwidth}
    \centering
  \includegraphics[width=1\textwidth]{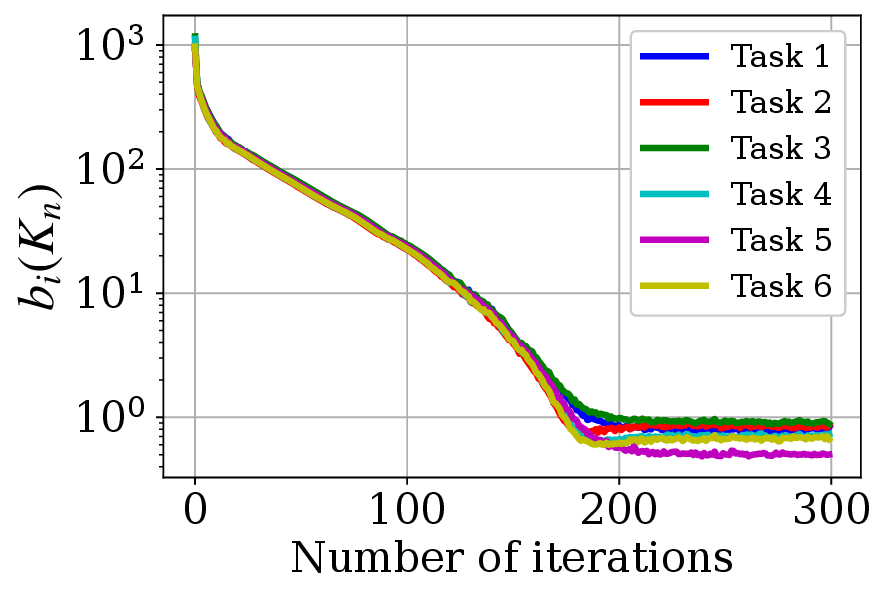}
  \end{minipage}
  \begin{minipage}{.32\textwidth}
    \centering    \includegraphics[width=1\textwidth]{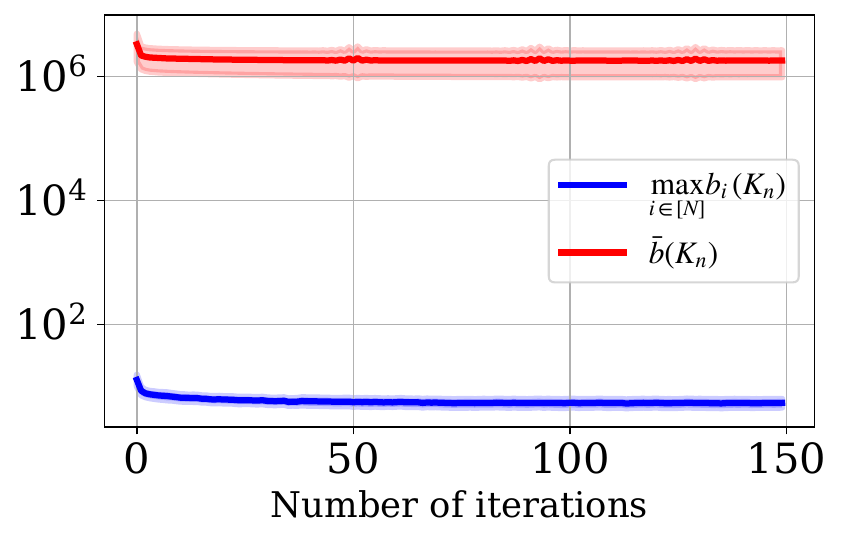}
  \end{minipage}
  \vspace{-0.2cm}
  \caption{Evolution of optimality gaps (left), bisimulation-based measures from \eqref{eq:avg_bis_het} (middle), and cost gradient discrepancy bounds (right) with respect to policy gradient iterations for multitask LQR with unicycle dynamics. On the right, we compare our bounds with those proposed in \cite{toso2024asynchronous, toso2024meta}.}
  \label{fig:unicycle}
\end{figure*}

Fig.~\ref{fig:unicycle} depicts the results for the multitask unicycle setting. On the left, we observe that the policy gradient controller achieves performance close to the task-specific optima, illustrating a setting favorable for collaborative control design.
In the middle, we note that our bisimulation-based measures are below $1$ at convergence, suggesting that the tasks are similar in closed loop with the designed common controller. This indicates the effectiveness of our task heterogeneity measure on capturing collaborative settings where the designed controller can be readily applied across distinct tasks. Further evaluation of our measures is reserved for future work.

On the right plot, we compare $\max_i b_i(K_n)$ with the measure $\bar{b}(K_n)$ from prior works~\cite{toso2024asynchronous, toso2024meta}, averaged over $75$ random collections of two unicycle tasks. We note that our measure is significantly less conservative, with a reduction of $99.9996\%$ at convergence, leading to non-vacuous task-specific optimality guarantees. In contrast, the previous heterogeneity measure is overly pessimistic and fails to capture the observed task similarities.

\section{Future Work}

Moving forward, our bisimulation-based measures could be applied to relax the task heterogeneity assumptions and enhance the performance bounds in related settings (e.g., LQR with domain randomization \cite{fujinami2025policy}, meta-learning LQR \cite{toso2024meta}). Moreover, we aim to leverage our new task heterogeneity measures to guide the learning of latent shared dynamics among multiple systems, providing a more favorable setting for applying multitask LQR. Another interesting direction is to employ our bisimulation-based measures in the design of policy gradient updates and provide robustness guarantees for multitask LQR design. Our results can also be extended for the analysis of the model-free setting.

\section{Acknowledgments}

We thank Bruce D. Lee for insightful discussions on
this work. Charis Stamouli and George J. Pappas acknowledge support from NSF award SLES-2331880. Leonardo F. Toso is funded by the CAIRFI and Presidential Fellowships. James Anderson is partially funded by NSF grants ECCS 2144634 and 2231350 and the Columbia DSI.

\bibliographystyle{IEEEtran} % use IEEEtran.bst style
\bibliography{arxiv_version}

\appendix
\bigskip
\medskip
\noindent\textbf{\Large Appendix}
\section{Proofs}\label{app:Proofs}
\subsection{Proof of Lemma~\ref{lem:asympt_V_bound}}
Let $\lambda\in(0,1)$ be such that \eqref{eq:BF_condition_2} is satisfied. 
Then, inequality \eqref{eq:BF_condition_2} can be written as:
\begin{equation}\label{eq:lem1_proof_1}
 V(\Sigma_{K,t+1}^{(i)},\Sigma_{K,t+1}^{(j)})\leq(1-\lambda) V(\Sigma_{K,t}^{(i)},\Sigma_{K,t}^{(j)})+V(\Sigma_0^{(i)},\Sigma_0^{(j)}).
\end{equation}
By recursively applying \eqref{eq:lem1_proof_1}, we obtain:
\begin{align}\label{eq:lem1_proof_2}
    V(\Sigma_{K,t+1}^{(i)},\Sigma_{K,t+1}^{(j)})&\leq(1-\lambda)^tV(\Sigma_0^{(i)},\Sigma_0^{(j)})+\sum_{\tau=0}^{t-1}(1-\lambda)^\tau V(\Sigma_0^{(i)},\Sigma_0^{(j)})\nonumber\\
    &=(1-\lambda)^tV(\Sigma_0^{(i)},\Sigma_0^{(j)})+\frac{1-(1-\lambda)^t}{\lambda}V(\Sigma_0^{(i)},\Sigma_0^{(j)}).
\end{align}
Since $\lambda\in(0,1)$, \eqref{eq:lem1_proof_2} implies that:
\begin{equation}\label{eq:lem1_proof_3}
\limsup_{t\to\infty}V(\Sigma_{K,t+1}^{(i)},\Sigma_{K,t+1}^{(j)})\leq V(\Sigma_0^{(i)},\Sigma_0^{(j)})/\lambda.
\end{equation}
As explained in the derivation of \eqref{eq:grad_het_limit}, we know that the limit $\lim_{t\to\infty}\norm{Y_{K,t}^{(i)}-Y_{K,t}^{(j)}}_F$ exists. Therefore, we have:
\begin{align*}
    \lim_{t\to\infty}\norm{Y_{K,t}^{(i)}-Y_{K,t}^{(j)}}_F &= \lim_{t\to\infty}\norm{E_K^{(i)} \Sigma_{K,t}^{(i)}-E_K^{(j)} \Sigma_{K,t}^{(j)}}_F \hspace*{1.1cm}\text{(from \eqref{eq:matrix_system})}\nonumber\\
    &\leq \limsup_{t\to\infty}V(\Sigma_{K,t}^{(i)},\Sigma_{K,t}^{(j)}) \hspace*{2.2cm}\text{(from \eqref{eq:BF_condition_1})}\nonumber\\
    &\leq V(\Sigma_0^{(i)},\Sigma_0^{(j)})/\lambda, \hspace*{2.9cm}\text{(from \eqref{eq:lem1_proof_3})}
\end{align*}
which completes the proof.

\subsection{Proof of Lemma~\ref{lem:M_K_conditions}}
We  prove the lemma by adapting the proof of \cite[Proposition 3.3]{girard2005} to discrete-time Lyapunov matrix systems, similar to the proof of \cite[Lemma 2]{stamouli2025layered} for discrete-time systems with vector states. Consider a fixed $K\in\calK_\stab$ and a pair of task indices $(i,j)$. Then, we know  that the eigenvalues of $A_K^{(ij)}$ lie within the unit disk. Let $M=\thickbar{M}+E_K^{(ij)\intercal}E_K^{(ij)}$, where $\thickbar{M}\in\setS_+^{2d_x}$ will be determined later on. Given that $\thickbar{M}$ is restricted within the set of positive definite matrices, we can directly deduce that $M$ is positive definite and inequality \eqref{eq:M_condition} is satisfied. We set:
\[
    A_{K,\lambda}^{(ij)} = \frac{1}{\sqrt{1-\lambda}} A_K^{(ij)},
\]
where $\lambda\in(0,1)$ is such that the eigenvalues of $A_{K,\lambda}^{(ij)}$ lie strictly within the unit disk. We note that we can always select a scalar $\lambda$ small enough such that this condition holds. By definition of $A_{K,\lambda}^{(ij)}$ and $M$, inequality \eqref{eq:constraint_stabilization} can be equivalently written as:
\begin{align}
    &A_K^{(ij)\intercal}MA_K^{(ij)}-(1-\lambda)M\preceq0 \nonumber\\
    \iff &A_{K,\lambda}^{(ij)\intercal}MA_{K,\lambda}^{(ij)}-M\preceq0\nonumber\\
    \label{lem:tracking_precision_guarantee_proof_2}
    \iff &A_{K,\lambda}^{(ij)\intercal}\thickbar{M}A_{K,\lambda}^{(ij)}-\thickbar{M}\preceq-A_{K,\lambda}^{(ij)\intercal}E_K^{(ij)\intercal}E_K^{(ij)}A_{K,\lambda}^{(ij)}+E_K^{(ij)\intercal}E_K^{(ij)}.
\end{align}
Let $\Lambda\succeq0$ be such that:
\begin{equation}\label{lem:tracking_precision_guarantee_proof_3}
    A_{K,\lambda}^{(ij)\intercal}E_K^{(ij)\intercal}E_K^{(ij)}A_{K,\lambda}^{(ij)}-E_K^{(ij)\intercal}E_K^{(ij)}\preceq \Lambda
\end{equation}
and define $\thickbar{M}$ as the unique positive definite solution of the Lyapunov equation:
\begin{equation}\label{lem:tracking_precision_guarantee_proof_4}
    A_{K,\lambda}^{(ij)\intercal}\thickbar{M}A_{K,\lambda}^{(ij)}-\thickbar{M}=-\Lambda.
\end{equation}
Substituting \eqref{lem:tracking_precision_guarantee_proof_4} into \eqref{lem:tracking_precision_guarantee_proof_3}, we obtain inequality \eqref{lem:tracking_precision_guarantee_proof_2}. Since \eqref{lem:tracking_precision_guarantee_proof_2} is equivalent to \eqref{eq:constraint_stabilization}, this completes the proof.

\subsection{Proof of Theorem~\ref{theorem:Bisimulation_Functions_characterization}}
Let us first note that for any $A,B\in\setS_+^{d_x}$, we have the basic result:
\begin{equation}\label{eq:thm1_proof_1}
    \lambda_{\min}(A)\trace{B}\leq\trace{AB}\leq\lambda_{\max}(A)\trace{B}.
\end{equation}
Fix any $\Sigma_{K,t}^{(i)},\Sigma_{K,t}^{(j)}\in\setS_+^{d_x}$. By using straightforward algebraic manipulations, we can write:
\begin{align}
    \norm{E_K^{(i)} \Sigma_{K,t}^{(i)}-E_K^{(j)} \Sigma_{K,t}^{(j)}}_F &= \sqrt{\trace{(E_K^{(i)} \Sigma_{K,t}^{(i)}-E_K^{(j)} \Sigma_{K,t}^{(j)})^\intercal(E_K^{(i)} \Sigma_{K,t}^{(i)}-E_K^{(j)} \Sigma_{K,t}^{(j)})}}\nonumber\\
    &=\sqrt{\trace{\vect{\Sigma_{K,t}^{(i)}}{\Sigma_{K,t}^{(j)}}E_K^{(ij)\intercal}E_K^{(ij)}\vectt{\Sigma_{K,t}^{(i)}}{\Sigma_{K,t}^{(j)}}}}\nonumber\\
    &=\sqrt{\trace{\vect{\mathds{I}_{d_x}}{\mathds{I}_{d_x}}\mydiag(\Sigma_{K,t}^{(i)},\Sigma_{K,t}^{(j)})E_K^{(ij)\intercal}E_K^{(ij)}\mydiag(\Sigma_{K,t}^{(i)},\Sigma_{K,t}^{(j)})\vectt{\mathds{I}_{d_x}}{\mathds{I}_{d_x}}}}\nonumber\\
    &=\sqrt{\trace{\begin{bmatrix}
        \mathds{I}_{d_x} & \mathds{I}_{d_x} \\
        \mathds{I}_{d_x} & \mathds{I}_{d_x}
    \end{bmatrix}\mydiag(\Sigma_{K,t}^{(i)},\Sigma_{K,t}^{(j)})E_K^{(ij)\intercal}E_K^{(ij)}\mydiag(\Sigma_{K,t}^{(i)},\Sigma_{K,t}^{(j)})}}\nonumber
\end{align}
\begin{align}
\label{eq:thm1_proof_2}&\leq\sqrt{2\trace{\mydiag(\Sigma_{K,t}^{(i)},\Sigma_{K,t}^{(j)})E_K^{(ij)\intercal}E_K^{(ij)}\mydiag(\Sigma_{K,t}^{(i)},\Sigma_{K,t}^{(j)})}},
\end{align}
where the last step follows from property \eqref{eq:thm1_proof_1} and the fact that: 
\[
\lambda_{\max}\left(\begin{bmatrix}
        \mathds{I}_{d_x} & \mathds{I}_{d_x} \\
        \mathds{I}_{d_x} & \mathds{I}_{d_x}
    \end{bmatrix}\right)=2,
\]
for any $d_x\in\setN_+$. Employing \eqref{eq:M_condition} and \eqref{eq:thm1_proof_2}, we get:
\begin{align}\label{eq:thm1_proof_3}
    \norm{E_K^{(i)} \Sigma_{K,t}^{(i)}-E_K^{(j)} \Sigma_{K,t}^{(j)}}_F &\leq \sqrt{2\trace{\mydiag(\Sigma_{K,t}^{(i)},\Sigma_{K,t}^{(j)})M\mydiag(\Sigma_{K,t}^{(i)},\Sigma_{K,t}^{(j)})}} \nonumber\\
    &\leq \sqrt{\frac{2}{\lambda_{\min}(M)}\trace{M\mydiag(\Sigma_{K,t}^{(i)},\Sigma_{K,t}^{(j)})M\mydiag(\Sigma_{K,t}^{(i)},\Sigma_{K,t}^{(j)})}} \hspace{0.5cm}\text{(from \eqref{eq:thm1_proof_1})}\nonumber\\
    &= \sqrt{\frac{2}{\lambda_{\min}(M)}\trace{Z_{K,t}^{(ij)2}}},
\end{align}
where $Z_{K,t}^{(ij)} = M^{1/2}\mydiag(\Sigma_{K,t}^{(i)},\Sigma_{K,t}^{(j)})M^{1/2}$.
Note that $Z_{K,t}^{(ij)}\in\setS_+^{2d_x}$, given that $\Sigma_{K,t}^{(i)}$, $\Sigma_{K,t}^{(j)}\in\setS_+^{2d_x}$. Therefore, by definition of trace, it is straightforward to show that:
\begin{equation}\label{eq:thm1_proof_4}
    \trace{Z_{K,t}^{(ij)}}\geq\sqrt{\trace{Z_{K,t}^{(ij)2}}}.
\end{equation}
Combining \eqref{eq:thm1_proof_3} and \eqref{eq:thm1_proof_4}, we deduce that the function \eqref{eq:bisimulation_function} satisfies \eqref{eq:BF_condition_1}.

Next, we show that the function \eqref{eq:bisimulation_function} satisfies inequality \eqref{eq:BF_condition_2}. Using \eqref{eq:bisimulation_function} and \eqref{eq:matrix_system}, we have:
\begin{align}
     &\hspace*{-4cm}V(\Sigma_{K,t+1}^{(i)},\Sigma_{K,t+1}^{(j)})- V(\Sigma_{K,t}^{(i)},\Sigma_{K,t}^{(j)}) \nonumber\\
     &\hspace*{-4cm}=\sqrt{\frac{2}{\lambda_{\min}(M)}}\left(\trace{M\mydiag(\Sigma_{K,t+1}^{(i)},\Sigma_{K,t+1}^{(j)})}-\trace{M\mydiag(\Sigma_{K,t}^{(i)},\Sigma_{K,t}^{(j)})}\right) \nonumber
\end{align}
\begin{align}
     &=\sqrt{\frac{2}{\lambda_{\min}(M)}}\left(\trace{M(A_K^{(ij)}\mydiag(\Sigma_{K,t}^{(i)},\Sigma_{K,t}^{(j)})A_K^{(ij)\intercal}+\mydiag(\Sigma_0^{(i)},\Sigma_0^{(j)}))}-\trace{M\mydiag(\Sigma_{K,t}^{(i)},\Sigma_{K,t}^{(j)})}\right) \nonumber\\
     &=\sqrt{\frac{2}{\lambda_{\min}(M)}}\left(\trace{(A_K^{(ij)\intercal}MA_K^{(ij)}-M)\mydiag(\Sigma_{K,t}^{(i)},\Sigma_{K,t}^{(j)})}+\trace{M\mydiag(\Sigma_0^{(i)},\Sigma_0^{(j)})}\right) \nonumber\\
     &\leq\sqrt{\frac{2}{\lambda_{\min}(M)}}\left(-\lambda\trace{M\mydiag(\Sigma_{K,t}^{(i)},\Sigma_{K,t}^{(j)})}+\trace{M\mydiag(\Sigma_0^{(i)},\Sigma_0^{(j)})}\right) \hspace{1cm}\text{(from \eqref{eq:constraint_stabilization})} \nonumber\\
     &=-\lambda V(\Sigma_{K,t}^{(i)},\Sigma_{K,t}^{(j)})+V(\Sigma_0^{(i)},\Sigma_0^{(j)}). \nonumber
\end{align}
Hence, we have shown condition \eqref{eq:BF_condition_2} of~\Cref{def:bisim_function} is also satisfied, which completes the proof.

\subsection{Proof of Theorem~\ref{thm:algor_indep_bounds}}
We prove the theorem by adapting the proof of \cite[Theorem 3]{toso2024meta}, developed for the setting of meta-learning LQR, to our setting of multitask LQR. Employing \cite[Lemma 11]{fazel2018global}, we can write:
\begin{align}\label{eq:thm2_proof_1}
    J^{(i)}(K_\star)-J^{(i)}(K_\star^{(i)})&\leq \frac{\big\|\Sigma_{K_\star^{(i)}}^{(i)}\big\|\trace{\nabla J^{(i)}(K_\star)^{\intercal}\nabla J^{(i)}(K_\star)}}{\lambda_{\min}(\Sigma_0^{(i)})^2\,\sigma_{\min}(R^{(i)})}.
\end{align}
By using \cite[Lemma A.1]{stamouli2024rate} and triangle inequality, we obtain:
\begin{align}\label{eq:thm2_proof_2}
    \trace{\nabla J^{(i)}(K_\star)^{\intercal}\nabla J^{(i)}(K_\star)} &= \norm{\nabla J^{(i)}(K_\star)}_F^2 \nonumber\\
    &\leq 2\norm{\nabla J_\avg(K_\star)}_F^2+2(\norm{\nabla J^{(i)}(K_\star)}_F-\norm{\nabla J_\avg(K_\star)}_F)^2 \nonumber\\
    &\leq 2\norm{\nabla J_\avg(K_\star)}_F^2+2\norm{\nabla J^{(i)}(K_\star)-\nabla J_\avg(K_\star)}_F^2 \nonumber\\
    &=2\norm{\nabla J^{(i)}(K_\star)-\nabla J_\avg(K_\star)}_F^2,
\end{align}
where the last step follows from the fact that $\nabla J_\avg(K_\star)=0$, since $K_\star$ is assumed to be a minimizer of \eqref{eq:average_cost_function}. By direct algebraic manipulations, \eqref{eq:thm2_proof_2} yields:
\begin{align}\label{eq:thm2_proof_3}
    \trace{\nabla J^{(i)}(K_\star)^{\intercal}\nabla J^{(i)}(K_\star)} &\leq 2\Bigg\|\frac{1}{N}\sum_{j=1}^N(\nabla J^{(i)}(K_\star)-\nabla J^{(j)}(K_\star))\Bigg\|_F^2 \nonumber\\
    &= 2\Bigg\|\frac{1}{N}\sum_{j\neq i}(\nabla J^{(i)}(K_\star)-\nabla J^{(j)}(K_\star))\Bigg\|_F^2 \nonumber\\
    &\leq 2\left(\frac{1}{N}\sum_{j\neq i}\norm{\nabla J^{(i)}(K_\star)-\nabla J^{(j)}(K_\star)}_F\right)^2 \hspace{0.3cm}\text{(from triangle inequality)} \nonumber\\
    &\leq 2\left(\frac{1}{N}\sum_{j\neq i}b_{ij}(K_\star)\right)^2 \hspace{4.2cm}\text{(from~\Cref{def:bis_task_het})} \nonumber\\
    &=2b_i(K_\star)^2,
\end{align}
where the last step follows from \eqref{eq:avg_bis_het}. We complete the proof by combining \eqref{eq:thm2_proof_1} and \eqref{eq:thm2_proof_2}.

\subsection{Proof of Theorem~\ref{thm:policy gradient_bounds}}

We begin with a novel lemma of independent interest, which generalizes \cite[Theorem 1.3]{hu2023toward} from single-task to multitask LQR. Our lemma provides conditions such that the policy gradient iterates defined by \eqref{eq:policy_gradient} remain stabilizing for all systems. We employ $\inner{\cdot}{\cdot}$ to denote the Euclidean inner product of two real vectors.

\begin{lemma}[Conditions for Stabilizing Multitask Policy Gradient LQR]\label{lem:policy_gradient_stability}
Let $K_0\in\calK_\stab$  and consider the sequence $\{K_n\}_{n=0}^\infty$ of policy gradient iterates given by \eqref{eq:policy_gradient}. Consider $\beta_1,\ldots,\beta_N\geq1$, and let $\calK$ denote the stabilizing subset from~\Cref{def:stabilizing_set}. For each $i$, let $L_i$ denote the smoothness constant of the $i$-th LQR cost function $J^{(i)}(\cdot)$ over the set $\calK$. For every task $\calT^{(i)}$ and any  $K\in\calK_\stab$, let $b_{i}(K)$ be given by \eqref{eq:avg_bis_het}.  Assume that $\alpha<8/\max_i\gamma_i,$ where  $\gamma_i=(4\lambda_{\min}(\Sigma_0^{(i)})^2\sigma_{\min}(R^{(i)}))/\norm{\Sigma_{K_\star^{(i)}}^{(i)}}$, for all $i=1,\ldots,N$. Moreover, assume that:
\begin{equation}\label{eq:heter_assumption}
    \sup_{K\in\calK}b_i(K)^2\leq \frac{\gamma_i}{6}(J^{(i)}(K_0)-J^{(i)}(K_\star^{(i)})).
\end{equation}
Then, the line segment connecting $K_n$ and $K_{n+1}$ is contained in $\calK$, for all $n\in\setN$.
\end{lemma}

\emph{Proof.} Let $K\in\calK$ and suppose we have chosen:
\begin{equation}\label{eq:lem3_proof_1}
    \textstyle  \alpha=8/(\max_i\gamma_i+\omega),
\end{equation}
for some constant $\omega>0$. Moreover, set $K'=K-\alpha\nabla J_{\avg}(K)$. Since the cost functions $J^{(i)}(\cdot)$ are twice continuously differentiable, we know that the functions $\norm{\nabla^2J^{(i)}(\cdot)}$ are continuous. Therefore, there exist small constants $c_i>0$ such that:
\begin{align}\label{eq:lem3_proof_2}
    \norm{\nabla^2J^{(i)}(\bar{K})}\leq L_i+\omega,\; i=1,\ldots,N, 
\end{align}
for all $\bar{K}\in\calK_c:=\cap_{i=1}^N\calK_c^{(i)}$, where:
\begin{align*}
    \calK_c^{(i)}=\Big\{\bar{K}\in\setR^{d_u\times d_x}&: J^{(i)}(\bar{K})-J^{(i)}(K_\star^{(i)})\leq\beta_i(J^{(i)}(K_0)-J^{(i)}(K_\star^{(i)}))+c_i\Big\}.
\end{align*}   
Let $\calS$ be the complement of the closure of $\calK_c$. Then, $\calK_c\cap\calS$ is empty, and since $\calK$ is compact, the distance between $\calK$ and $\calS$ is strictly positive. Let $\delta$ denote this distance and set:
\begin{equation}\label{eq:lem3_proof_3}
    \tau = \min\left\{(0.9\delta)/\norm{\nabla J_\avg(K)},1/(4(\max_j L_j+\omega)),4/(\max_j\gamma_j+\omega)\right\}.
\end{equation}
Fix any $\theta\in[0,1]$ and define the point $K_\theta:=(1-\theta)K+\theta(K-\tau\nabla J_{\avg}(K))$ contained in the line segment connecting $K$ and $K-\tau\nabla J_{\avg}(K)$. Then, we have:
\begin{equation}\label{eq:lem3_proof_4}
    K_\theta=K-\theta\tau\nabla J_\avg(K).
\end{equation}
From \eqref{eq:lem3_proof_3} and \eqref{eq:lem3_proof_4} we deduce that:
\begin{align*}
    \norm{K-K_\theta}=\norm{\theta\tau\nabla J_{\avg}(K)}\leq \tau\norm{\nabla J_{\avg}(K)}\leq\frac{0.9\delta}{\norm{\nabla J_{\avg}(K)}}\norm{\nabla J_{\avg}(K)}=0.9\delta.
\end{align*}
Hence, we can conclude that the line segment connecting $K$ and $K-\tau\nabla J_\avg(K)$ is contained in $\calK_c$. Fix any $i\in\{1,\ldots,N\}$. Since $\norm{\nabla^2J^{(i)}(\bar{K})}\leq L_i+\omega$, for all $\bar{K}\in\calK_c$, we have:
\begin{align}\label{eq:lem3_proof_5}
    J^{(i)}(K&-\tau\nabla J_\avg(K))-J^{(i)}(K)\nonumber\\
    &\leq\; \inner{\nabla J^{(i)}(K)}{K-\tau\nabla J_\avg(K)-K}+\frac{L_i+\omega}{2}\norm{K-\tau\nabla J_\avg(K)-K}_F^2 \nonumber\\
    &=\inner{\nabla J^{(i)}(K)}{-\tau\nabla J_\avg(K)}+\frac{(L_i+\omega)\tau^2}{2}\norm{\nabla J_\avg(K)}_F^2\nonumber\\
    &=\inner{\nabla J^{(i)}(K)}{-\tau\nabla J^{(i)}(K)+\tau\nabla J^{(i)}(K)-\tau\nabla J_\avg(K)}+\frac{(L_i+\omega)\tau^2}{2}\norm{\nabla J_\avg(K)}_F^2\nonumber\\
    &\leq -\tau\norm{\nabla J^{(i)}(K)}_F^2+\tau\inner{\nabla J^{(i)}(K)}{\nabla J^{(i)}(K)-\nabla J_\avg(K)}\nonumber\\
    &\hspace*{0.45cm}+\frac{\tau}{8}\norm{-\nabla J^{(i)}(K)+\nabla J^{(i)}(K)-\nabla J_\avg(K)}_F^2,
\end{align}
where the last inequality follows from \eqref{eq:lem3_proof_3}. By applying Young's inequality to the second term and \cite[Lemma A.1]{stamouli2024rate} to the third term on the right-hand side of \eqref{eq:lem3_proof_5}, we obtain:
\begin{align}\label{eq:lem3_proof_6}
    J^{(i)}(K&-\tau\nabla J_\avg(K))-J^{(i)}(K)\nonumber\\
    &\leq -\tau\norm{\nabla J^{(i)}(K)}_F^2+\frac{\tau}{2}\norm{\nabla J^{(i)}(K)}_F^2+\frac{\tau}{2}\norm{\nabla J^{(i)}(K)-\nabla J_\avg(K)}_F^2\nonumber\\
    &\hspace{0.45cm}+\frac{\tau}{4}\norm{\nabla J^{(i)}(K)}_F^2+\frac{\tau}{4}\norm{\nabla J^{(i)}(K)-\nabla J_\avg(K)}_F^2\nonumber\\
    &=-\frac{\tau}{4}\norm{\nabla J^{(i)}(K)}_F^2+\frac{3\tau}{4}\norm{\nabla J^{(i)}(K)-\nabla J_\avg(K)}_F^2.
\end{align}
From \cite[Lemma 3]{toso2024meta} we know that:
\begin{equation}\label{eq:lem3_proof_7}
    \norm{\nabla J^{(i)}(K)}_F^2\geq\gamma_i(J^{(i)}(K)-J^{(i)}(K_{\star}^{(i)})).
\end{equation}
Similar to \eqref{eq:thm2_proof_3}, we can show that:
\begin{equation*}
    \norm{\nabla J^{(i)}(K)-\nabla J_\avg(K)}_F^2\leq b_i(K)^2,
\end{equation*} 
and \eqref{eq:heter_assumption} yields:
\begin{equation}\label{eq:lem3_proof_8}
    \norm{\nabla J^{(i)}(K)-\nabla J_\avg(K)}_F^2\leq \frac{\gamma_i}{6}(J^{(i)}(K_0)-J^{(i)}(K_\star^{(i)})). 
\end{equation}
Employing \eqref{eq:lem3_proof_7} and \eqref{eq:lem3_proof_8}, inequality \eqref{eq:lem3_proof_6} implies that:
\begin{align*}
    &J^{(i)}(K-\tau\nabla J_\avg(K))-J^{(i)}(K)\leq -\frac{\tau\gamma_i}{4}(J^{(i)}(K)-J^{(i)}(K_{\star}^{(i)}))+\frac{\tau\gamma_i}{8}(J^{(i)}(K_0)-J^{(i)}(K_\star^{(i)})),
\end{align*}
or equivalently:
\begin{align}\label{eq:lem3_proof_9}
    &J^{(i)}(K-\tau\nabla J_\avg(K))-J^{(i)}(K_\star^{(i)})\leq\left(1-\frac{\tau\gamma_i}{4}\right)(J^{(i)}(K)-J^{(i)}(K_{\star}^{(i)}))+\frac{\tau\gamma_i}{8}(J^{(i)}(K_0)-J^{(i)}(K_\star^{(i)})).
\end{align}
Since $K\in\calK$, by~\Cref{def:stabilizing_set} we obtain:
\begin{equation}\label{eq:lem3_proof_10}
    J^{(i)}(K)-J^{(i)}(K_{\star}^{(i)})\leq \beta_i(J^{(i)}(K_0)-J^{(i)}(K_{\star}^{(i)})).
\end{equation}
Given \eqref{eq:lem3_proof_10} and the fact that $\tau<4/\gamma_i$ (from \eqref{eq:lem3_proof_3}) and $\beta_1,\ldots,\beta_N\geq1$, \eqref{eq:lem3_proof_9} implies that:
\begin{align*}
     J^{(i)}(K&-\tau\nabla J_\avg(K))-J^{(i)}(K_\star^{(i)})\\
     &\leq \left(1-\frac{\tau\gamma_i}{4}\right)\beta_i(J^{(i)}(K_0)-J^{(i)}(K_{\star}^{(i)}))+\frac{\tau\gamma_i\beta_i}{8}(J^{(i)}(K_0)-J^{(i)}(K_\star^{(i)}))\nonumber\\
     &=\left(1-\frac{\tau\gamma_i}{8}\right)\beta_i(J^{(i)}(K_0)-J^{(i)}(K_{\star}^{(i)}))\nonumber\\
     &\leq \beta_i(J^{(i)}(K_0)-J^{(i)}(K_{\star}^{(i)})),
\end{align*}
and hence we have $K-\tau\nabla J_\avg(K)\in\calK$.
In fact, it is straightforward to see that the line segment connecting $K$ and $K-\tau\nabla J_\avg(K)$ is contained in $\calK$ by varying $\tau$. By applying the same procedure, we can show that the line segment connecting $K-\tau\nabla J_\avg(K)$ and $K-2\tau\nabla J_\avg(K)$ is contained in $\calK$. Since $\tau>0$, we only need to apply this procedure $\ceil{\alpha/\tau}$ times to show that the line segment between $K$ and $K-\alpha\nabla J_\avg(K)$ is in $\calK$.

Now set $K=K_0$. Based on our proof above and the fact that $K_0\in\calK$ (by definition of $\calK$), we can conclude that the line segment between $K_0$ and $K_1:=K_0-\alpha\nabla J_\avg(K_0)$ is in $\calK$. Using the same procedure, we can sequentially show that the line segments between $K_1$ and $K_2$, $K_2$ and $K_3$, \ldots, are in $\calK$, which completes the proof.

$\qed$

Fix any $i\in\{1,\ldots,N\}$. Based on the above lemma and the fact that $J^{(i)}(\cdot)$ is $L_i$-smooth on $\calK$, we can deduce that:
\begin{align*}%\label{eq:thm3_proof_1}
 \norm{\nabla^2 J^{(i)}(K)}\leq L_i,    
\end{align*}
for all $K$ in the line segment connecting $K_{n-1}$ and $K_n$, for all $n\in\setN_+$. Hence, we can write:
\begin{align}\label{eq:thm3_proof_2}
    J^{(i)}&(K_n)-J^{(i)}(K_{n-1})\nonumber\\
    &\leq\; \inner{\nabla J^{(i)}(K_{n-1})}{K_n-K_{n-1}}+\frac{L_i}{2}\norm{K_n-K_{n-1}}_F^2 \nonumber\\
    &=\inner{\nabla J^{(i)}(K_{n-1})}{-\alpha\nabla J_\avg(K_{n-1})}+\frac{L_i\alpha^2}{2}\norm{\nabla J_\avg(K_{n-1})}_F^2\nonumber\\
    &=\inner{\nabla J^{(i)}(K_{n-1})}{-\alpha\nabla J^{(i)}(K_{n-1})+\alpha\nabla J^{(i)}(K_{n-1})-\alpha\nabla J_\avg(K_{n-1})}+\frac{L_i\alpha^2}{2}\norm{\nabla J_\avg(K_{n-1})}_F^2\nonumber\\
    &\leq -\alpha\norm{\nabla J^{(i)}(K_{n-1})}_F^2+\alpha\inner{\nabla J^{(i)}(K_{n-1})}{\nabla J^{(i)}(K_{n-1})-\nabla J_\avg(K_{n-1})}\nonumber\\
    &\hspace*{0.45cm}+\frac{\alpha}{8}\norm{-\nabla J^{(i)}(K_{n-1})+\nabla J^{(i)}(K_{n-1})-\nabla J_\avg(K_{n-1})}_F^2,
\end{align}
where the last inequality follows from the assumption that $\alpha<1/4L_i$. By applying Young's inequality to the second term and \cite[Lemma A.1]{stamouli2024rate} to the third term on the right-hand side of \eqref{eq:thm3_proof_2}, we obtain:
\begin{align}\label{eq:thm3_proof_3}
    J^{(i)}(K_n)&-J^{(i)}(K_{n-1})\nonumber\\
    &\leq -\alpha\norm{\nabla J^{(i)}(K_{n-1})}_F^2+\frac{\alpha}{2}\norm{\nabla J^{(i)}(K_{n-1})}_F^2+\frac{\alpha}{2}\norm{\nabla J^{(i)}(K_{n-1})-\nabla J_\avg(K_{n-1})}_F^2\nonumber\\
    &\hspace{0.45cm}+\frac{\alpha}{4}\norm{\nabla J^{(i)}(K_{n-1})}_F^2+\frac{\alpha}{4}\norm{\nabla J^{(i)}(K_{n-1})-\nabla J_\avg(K_{n-1})}_F^2\nonumber\\
    &=-\frac{\alpha}{4}\norm{\nabla J^{(i)}(K_{n-1})}_F^2+\frac{3\alpha}{4}\norm{\nabla J^{(i)}(K_{n-1})-\nabla J_\avg(K_{n-1})}_F^2.
\end{align}
Similar to \eqref{eq:thm2_proof_3}, we can show that for every $K\in\calK$:
\begin{equation*}
    \norm{\nabla J^{(i)}(K)-\nabla J_\avg(K)}_F^2\leq b_i(K)^2,
\end{equation*}
and Lemma~\ref{lem:policy_gradient_stability} yields:
\begin{equation}\label{eq:thm3_proof_4}
    \norm{\nabla J^{(i)}(K_{n-1})-\nabla J_\avg(K_{n-1})}_F^2\leq b_i(K_{n-1})^2.
\end{equation}
From \cite[Lemma 3]{toso2024meta}, \eqref{eq:thm3_proof_3}, and \eqref{eq:thm3_proof_4} we can conclude that:
\begin{align*}
    &J^{(i)}(K_n)-J^{(i)}(K_{n-1})\leq -\frac{\alpha\gamma_i}{4}(J^{(i)}(K_{n-1})-J^{(i)}(K_{\star}^{(i)}))+\frac{3\alpha}{4}b_i(K_{n-1})^2,
\end{align*}
or equivalently:
\begin{align}\label{eq:thm3_proof_5}
    &J^{(i)}(K_n)-J^{(i)}(K_\star^{(i)})\leq\left(1-\frac{\alpha\gamma_i}{4}\right)(J^{(i)}(K_{n-1})-J^{(i)}(K_{\star}^{(i)}))+\frac{3\alpha}{4}b_i(K_{n-1})^2.
\end{align}
Given the assumption that $\alpha<4/\gamma_i$, by recursively applying \eqref{eq:thm3_proof_5}, we obtain:
\begin{align}\label{eq:thm3_proof_6}
    J^{(i)}(K_n)-J^{(i)}(K_\star^{(i)})\leq\left(1-\frac{\alpha\gamma_i}{4}\right)^n(J^{(i)}(K_0)-J^{(i)}(K_{\star}^{(i)}))+\frac{3\alpha}{4}\sum_{\ell=0}^{n-1}\left(1-\frac{\alpha\gamma_i}{4}\right)^{n-\ell-1}b_i(K_{\ell})^2.
\end{align}
Set: $$S_n=\sum_{\ell=0}^{n-1}\left(1-\frac{\alpha\gamma_i}{4}\right)^{n-\ell-1}b_i(K_{\ell})^2.$$ Then, we have:
\begin{align*}
    S_{n+1} &= \left(1-\frac{\alpha\gamma_i}{4}\right)\sum_{\ell=0}^{n}\left(1-\frac{\alpha\gamma_i}{4}\right)^{n-\ell-1}b_i(K_\ell)^2\\
    &= \left(1-\frac{\alpha\gamma_i}{4}\right)\left(S_n+\left(1-\frac{\alpha\gamma_i}{4}\right)^{-1}b_i(K_n)^2\right)\\
    &=\left(1-\frac{\alpha\gamma_i}{4}\right) S_n+b_i(K_n)^2.
\end{align*}
By taking the limit superior on both sides of the above equality, we get:
\begin{align*}
    \limsup_{n\to\infty}S_{n+1}=\left(1-\frac{\alpha\gamma_i}{4}\right)\limsup_{n\to\infty}S_n+\limsup_{n\to\infty}b_i(K_n)^2\implies \limsup_{n\to\infty}S_n=\frac{4}{\alpha\gamma_i}\limsup_{n\to\infty}b_i(K_n)^2.
\end{align*}
Hence, since $1-\frac{\alpha\gamma_i}{4}\in(0,1)$, from \eqref{eq:thm3_proof_6} we deduce that:
\begin{align*}
    \limsup_{n\to\infty}(J^{(i)}(K_n)-J^{(i)}(K_\star^{(i)}))\leq \frac{3\alpha}{4}\limsup_{n\to\infty}S_n=\frac{3}{\gamma_i}\limsup_{n\to\infty}b_i(K_n)^2,
\end{align*}
which completes the proof.

\section{Computation of Bisimulation-Based Task Heterogeneity}\label{app:Computation of Bisimulation Measures of Task Heterogeneity}
In this subsection, we present a method of computing the quantities $M_K^{(ij)}$ and $\lambda_K^{(ij)}$ that appear in the bisimulation measure \eqref{eq:bisim_het}. Note that $M_K^{(ij)}$ and $\lambda_K^{(ij)}$ result from solving the following optimization problem:
\begin{subequations}\label{eq:bis_computation_proof_1}
\begin{align}
    \min_{\lambda\in(0,1),\, M\succ0}\;&\frac{\sqrt{2}\,\trace{M\mydiag(\Sigma_0^{(i)},\Sigma_0^{(j)})}}{\lambda\sqrt{\lambda_{\min}(M)}} \\
    \label{eq:bis_computation_proof_2}
    \subto &M\succeq E_K^{(ij)\intercal}E_K^{(ij)} \\
    \label{eq:bis_computation_proof_3}
    &A_K^{(ij)\intercal} M A_K^{(ij)}-M\preceq -\lambda M,
\end{align}
\end{subequations}
where the cost corresponds to the bound in \eqref{eq:grad_het_bis_bound} and the constraints represent the conditions of Lemma~\ref{lem:M_K_conditions}. 
We start by fixing $M$ and optimizing only with respect to $\lambda$. Note that the cost of \eqref{eq:bis_computation_proof_1} decreases for larger values of $\lambda$. On the other hand, as explained in the proof of~\Cref{lem:M_K_conditions}, \eqref{eq:bis_computation_proof_3} is feasible if and only if all the eigenvalues of:
\[
    A_{K,\lambda}^{(ij)} := \frac{1}{\sqrt{1-\lambda}} A_K^{(ij)}
\]
lie within the unit disk, that is:
\begin{equation}
    \rho(A_{K,\lambda}^{(ij)})<1 \iff \frac{\rho(A_K^{(ij)})}{\sqrt{1-\lambda}}<1 \iff \lambda<1-\rho(A_K^{(ij)})^2.
\end{equation}
Since $A_K^{(ij)}$ is (asymptotically) stable, we have $1-\rho(A_K^{(ij)})^2\in(0,1)$, and thus we can set:
\begin{equation}
    \lambda_K^{(ij)}=1-\rho(A_K^{(ij)})^2-\varepsilon,
\end{equation}
for some small constant $\varepsilon>0$. We note in passing that the introduction of a small constant  $\varepsilon$ is standard when addressing optimization problems with strict inequality constraints.

Next, we fix $\lambda=\lambda_K^{(ij)}$ and solve \eqref{eq:bis_computation_proof_1} with respect to $M$. In particular, we consider the problem:
\begin{subequations}\label{eq:bis_computation_proof_4}
\begin{align}
    \min_{\, M\succeq\varepsilon\mathds{I}_{2d_x}}\;&\frac{\sqrt{2}\,\trace{M\mydiag(\Sigma_0^{(i)},\Sigma_0^{(j)})}}{\lambda\sqrt{\lambda_{\min}(M)}} \\
    \subto\;\;\; &M\succeq E_K^{(ij)\intercal}E_K^{(ij)} \\
    &A_K^{(ij)\intercal} M A_K^{(ij)}-M\preceq -\lambda M,
\end{align}
\end{subequations}
where $\varepsilon$ is a small positive constant. We define a new optimization variable $s\geq\varepsilon$ and consider the equivalent problem:
\begin{subequations}\label{eq:bis_computation_proof_5}
\begin{align*}
    \min_{\, M,\,s\geq\varepsilon}\;&\frac{\sqrt{2}\,\trace{M\mydiag(\Sigma_0^{(i)},\Sigma_0^{(j)})}}{\lambda\sqrt{s}} \\
    \subto\;\;\; &s\leq\lambda_{\min}(M) \\
    &M\succeq E_K^{(ij)\intercal}E_K^{(ij)} \\
    &A_K^{(ij)\intercal} M A_K^{(ij)}-M\preceq -\lambda M,
\end{align*}
\end{subequations}
which can be written as:
\begin{subequations}\label{eq:bis_computation_proof_6}
\begin{align}
    \min_{\, M,\,s}\;\;&\frac{\sqrt{2}\,\trace{M\mydiag(\Sigma_0^{(i)},\Sigma_0^{(j)})}}{\lambda\sqrt{s}} \\
    \subto\;\, &M\succeq s\mathds{I}_{2d_x}, s\geq\varepsilon \\
    &M\succeq E_K^{(ij)\intercal}E_K^{(ij)} \\
    &A_K^{(ij)\intercal} M A_K^{(ij)}-M\preceq -\lambda M.
\end{align}
\end{subequations}
The epigraph form \cite{boyd2004convex} of \eqref{eq:bis_computation_proof_6} is given by:
\begin{subequations}\label{eq:bis_computation_proof_7}
\begin{align}
    \min_{\, M,\,s, u}\;& 2\sqrt{u} \\
    \subto\;\;\; 
    \label{eq:bis_computation_proof_8}&\frac{\sqrt{2}\,\trace{M\mydiag(\Sigma_0^{(i)},\Sigma_0^{(j)})}}{\lambda\sqrt{s}}\leq2\sqrt{u}\\
     &M\succeq s\mathds{I}_{2d_x}, s\geq\varepsilon \\
     &M\succeq E_K^{(ij)\intercal}E_K^{(ij)} \\
    &A_K^{(ij)\intercal} M A_K^{(ij)}-M\preceq -\lambda M.
\end{align}
\end{subequations}
Note that \eqref{eq:bis_computation_proof_8} can be expressed as a second-order cone constraint as follows:
\begin{align}
    \frac{\sqrt{2}\,\trace{M\mydiag(\Sigma_0^{(i)},\Sigma_0^{(j)})}}{\lambda}\leq 2\sqrt{us} \iff \left\|
\begin{bmatrix}
\displaystyle \left(\sqrt{2}\,\trace{M\mydiag(\Sigma_0^{(i)},\Sigma_0^{(j)})}\right)/\lambda \\[4pt]
u - s
\end{bmatrix}
\right\| \le u + s.
\end{align}
Therefore, the optimal value $M_K^{(ij)}$ can be computed via the following convex problem:
\begin{subequations}
\begin{align*}
    \min_{\, M,\,s, u}\;\;& 2\sqrt{u} \\
    \subto\;\, 
    &\left\|
\begin{bmatrix}
\displaystyle \left(\sqrt{2}\,\trace{M\mydiag(\Sigma_0^{(i)},\Sigma_0^{(j)})}\right)/\lambda \\[4pt]
u - s
\end{bmatrix}
\right\| \le u + s\\
     &M\succeq s\mathds{I}_{2d_x}, s\geq\varepsilon \\
     &M\succeq E_K^{(ij)\intercal}E_K^{(ij)} \\
    &A_K^{(ij)\intercal} M A_K^{(ij)}-M\preceq -\lambda M
\end{align*}
\end{subequations}
using standard solvers, such as those in CVXPY \cite{diamond2016cvxpy}.

\end{document}